\documentclass[eqsecnum,showkeys,noshowpacs,nofootinbib,amssymb,aps,epsfig]{revtex4}
\renewcommand{\theequation}{\arabic{equation}}
\def\beq{\begin{equation}}
\def\eeq{\end{equation}}
\def\bea{\begin{eqnarray}}
\def\eea{\end{eqnarray}}
\def\nn{\nonumber}

\def\pa{\partial}

\def\na{\nabla}

\begin{document}

\title{Dirac type relativistic quantum mechanics for massive photons}
\author{Soon-Tae Hong}
\email{galaxy.mass@gmail.com}
\affiliation{Center for Quantum Spacetime and Department of Physics, Sogang University, Seoul 04107, Korea}
\date{\today}
\begin{abstract}
Constructing a relativistic quantum mechanics (RQM) for a massive photon, without appealing to the quantum field theoretical approach such as the massive Proca model, we find a new theoretical particle solution which allows the massive photon having either positive or negative energy as solutions. In particular, we predict the existence of the so-called anti-photon corresponding to the negative energy solution, similar to the positron in the Dirac RQM for an electron. Note that 
the anti-photons could be an intense radiation flare of the gamma ray burst. In this RQM for the massive photon, 
we construct a positive definite probability density and a 
nontrivial diagonal Hamiltonian, and also discuss a massless photon.
Moreover we confirm the covariances of 
the relativistic equation of motion for the massive photon and the corresponding probability continuity equation under 
the Lorentz transformation.
\end{abstract}
\maketitle

\section{Introduction}
\setcounter{equation}{0}
\renewcommand{\theequation}{\arabic{section}.\arabic{equation}}

Generalizing the Hawking-Penrose singularity theorem (HPST)~\cite{hawking70}, the stringy cosmology in a higher dimensional total 
spacetime has been investigated~\cite{hong11,hong112} with a success that one can describe precisely 
the motion types of stringy congruence in terms of the universe expansion rate after the Big Bang.  Moreover, in the stringy HPST described in 
the higher dimensions, we can have an advantage that the degrees of freedom (DOF) of the rotation and shear of the stringy congruence are introduced naturally in the early universe. In the stringy HPST with the higher dimensional spacetime, assuming that the smallest particle is the photon, one can find that in the universe there exists a 
massive photon possessing the finite size which is filled with mass. Note that the higher dimensional theories have been exploited in the higher dimensional de Sitter cosmology~\cite{popov}. 

Next, the gamma ray burst (GRB) in the universe has been detected in 1967, and the GRB discovery has been published in 1973~\cite{strong73}. Since then, to explain the GRB, there have been extremely lots of theoretical models. For a recent detection of the GRB associated with the magnetar, see Ref.~\cite{hurley21} for instance. If the GRB sources were from within the Milky Way, they would be strongly concentrated near the galactic plane. The absence of any such pattern in the GRB has suggested a strong evidence that the GRB can come from beyond our own galaxy. The intense radiation flare of most detected GRB has been then assumed to emerge from a supernova, during a high-mass star 
implodes to construct a neutron star or a black hole.

On the other hand, Dirac has formulated the relativistic quantum mechanics (RQM) for an electron~\cite{dirac}. In his theory, a new particle solution has been proposed to allow the electron having 
either positive or negative energy solutions. In particular, the positron corresponding to the negative energy solution has 
been theoretically predicted and later has been experimentally confirmed~\cite{anderson33}. Note that the Dirac equation in condensed matter systems has been investigated in terms of bound state in continuum like solutions, in addition to discrete energy bound state solutions~\cite{panella12}. 

Recently making use of an open string which performs both rotational and pulsating motions, we have predicted the intrinsic frequency 
$\omega_{\gamma}=9.00\times 10^{23}~{\rm sec}^{-1}$~\cite{hong22} for the bosonic photon with spin one which is comparable to the intrinsic frequencies 
$\omega_{N}=0.87\times 10^{23}~{\rm sec}^{-1}$ and $\omega_{\Delta}=1.74\times 10^{23}~{\rm sec}^{-1}$~\cite{hong21} of the nucleon and delta baryon 
with spin 1/2, respectively. Next we have calculated the finite photon size $\langle r^{2}\rangle^{1/2}(\rm photon)=0.17~{\rm fm}$ in the phenomenological 
stringy photon model~\cite{hong22}. 

In this paper, we will propose a new RQM for a massive photon with finite size, without appealing to the quantum field theoretical 
approach such as the massive Proca model. In this RQM for the massive photon, we will theoretically predict a new particle solution which allows the massive photon having either positive or negative energy as solutions. In the massless limit we will also recover the RQM for the photon with transverse polarizations only. Moreover, we will investigate the intense radiation flare of the GRB in terms of the negative energy solution of the massive photon. Next we will discuss the covariances of 
the relativistic equation of motion for the massive photon and its corresponding probability continuity equation under 
the Lorentz transformation.

In Sec. II, we will construct the Hamiltonian for a massive photon in our model. In Sec. III, we will investigate 
the phenomenology of the RQM for the massive photon. Explicitly we will show that the probability density for the massive photon is 
positive definite in our model. The negative energy solution of the massive photon will be also discussed together with the GRB. 
In Sec. IV, we will study the massless limit 
of the photon. We will also compare the RQM for the massive photon, with the Proca model. 
In Sec. V, we will investigate the Lorentz transformation for the massive photon to discuss 
the covariances of its relativistic equation of motion and probability continuity equation. Sec. VI includes conclusions.

\section{Hamiltonian construction in RQM for a massive photon}
\setcounter{equation}{0}
\renewcommand{\theequation}{\arabic{section}.\arabic{equation}}

In this section, we will construct the Hamiltonian in the RQM for a massive photon. To do this, we assume 
that the photon trajectory is a straight line along 
the $z$ direction, for simplicity. We then have a relativistic relation $E^{2}=m^{2}+p^{2}$ where $p=p_{z}=|\vec{p}|$. 
The RQM equation of motion for the massive photon is then given by
\beq
H\phi^{a}=i\partial_{0}\phi^{a}.
\label{hpsi}
\eeq 
Here $\phi^{a}$ denotes the wave function for the massive photon, explicitly given by 
\beq
\phi^{a}_{A}=(\phi_{1}^{a},\phi_{2}^{a})^{t},
\label{phi2t}
\eeq
where the superscript $t$ stands for the transpose of the wave function components. Here the spin index $a$ $(a=0,1,2,3)$ 
denotes the spin DOF for the massive photon with spin one. The component index $A$ $(A=1,2)$ stands for the two DOF which have the same DOF of the positive and negative energy solutions with the energy index $\pm$ in (\ref{phipm}), since 
the positive and negative energy solutions are given by linear combinations of the two wave functions with the component 
indices. Note that the wave function $\phi^{a}_{A}$ is described in terms of a $1\times(2_{energy}\otimes 4_{spin})=1\times 8$ column vector. 
From now on we will drop the index $A$ in the wave functions except (\ref{vectorj}) and (\ref{boxm2}) below, for simplicity. 

Note that, in the quantum field theoretical Proca model for the massive photon, we need a $1\times 1$ matrix 
Hamiltonian as in (\ref{boxaa2}) below. Moreover, in this model, we cannot have a 
negative energy solution. Now we include a possibility of a negative energy solution for the massive photon in our model, by following 
Dirac idea for the positron. To do this, for the corresponding Hamiltonian for the bosonic massive photon, we introduce 
a minimal $2\times 2$ matrix associated with the positive and negative energy solutions.\footnote{This feature will be applied to the 
Lorentz covariance of the relativistic equation of motion for the massive photon in Sec. V. Note that, in the RQM for 
a fermionic electron, the wave function is described by $1\times (2_{energy}\otimes 2_{spin})=1\times 4$ column vector 
in (\ref{psixdirac})--(\ref{negsoldirac}) below originated from the positive and negative energy solutions in addition 
to $1/2$ and $-1/2$ spin states, while 
the corresponding Hamiltonian is given by a minimal $4\times 4$ matrix in (\ref{eomdirac}).}

Next, complying with the Dirac algorithm for the RQM for the positron, we proceed to find $H$ in (\ref{hpsi}) for the case of the RQM for the 
massive photon. The Hamiltonian $H$ is then given by a $2\times 2$ matrix acting on the component index $A$ only
\beq
H=\vec{\cal A}\cdot\vec{p}+{\cal B}m,
\label{hamiltonian}
\eeq
where ${\cal A}_{i}$ ($i=1,2,3$) and ${\cal B}$ are $2\times 2$ matrices. Using the relation $E^{2}=m^{2}+p^{2}$, we obtain the algebra among 
${\cal A}_{i}$ and ${\cal B}$ as follows
\beq
\{{\cal A}_{i},{\cal A}_{j}\}=2\delta_{ij}I,~~~{\cal B}^{2}=I,~~~\{{\cal A}_{i},{\cal B}\}=0,~~~
{\cal A}_{i}^{\dagger}={\cal A}_{i},~~~{\cal B}^{\dagger}={\cal B},
\label{aibeta}
\eeq
where $I$ is a $2\times 2$ unit matrix. Note that
eigenvalues of ${\cal A}_{i}$ or ${\cal B}$ are $\pm 1$ and ${\rm tr}{\cal A}_{i}={\rm tr}{\cal B}=0$. In our construction, the photon spin DOF is included in 
the wave function in (\ref{phi2t}), as in the Proca model wave function in (\ref{boxaa2}).

As in the Dirac relativistic formalism for the positron, exploiting the above relations in (\ref{aibeta}) together with 
the massive photon Hamiltonian in (\ref{hamiltonian}), we obtain the representations for ${\cal A}_{i}$ and ${\cal B}$ given by 
${\cal A}_{1}={\cal A}_{2}=0$, ${\cal A}_{3}=\sigma_{1}$ and ${\cal B}=\sigma_{3}$, with $\sigma_{i}$ being the Pauli matrices. Inserting the 
above representations for ${\cal A}_{i}$ and ${\cal B}$ into (\ref{hamiltonian}), we arrive at the desired $2\times 2$ Hamiltonian of the form
\beq
H=\left(
\begin{array}{cc}
m &-i\partial_{3}\\
-i\partial_{3} &-m 
\end{array}
\right).
\label{hpsi2}
\eeq
Making use of (\ref{hpsi}) and (\ref{hpsi2}), we find the relativistic equation of motion for the massive photon as follows
\beq 
(i\Gamma^{\mu}\pa_{\mu}-m)\phi^{a}(x)=0,
\label{brqm2}
\eeq
where $\phi^{a}(x)$ is a function of $x^{\mu}$ and $\Gamma^{\mu}$ is given by
\beq 
\Gamma^{\mu}\equiv (\Gamma^{0}, \Gamma^{1}, \Gamma^{2}, \Gamma^{3})=(\sigma_{3}, 0, 0, i\sigma_{2}).
\label{gammamu4}
\eeq

\section{Anti-photon in RQM for a massive photon}
\setcounter{equation}{0}
\renewcommand{\theequation}{\arabic{section}.\arabic{equation}}

In this section, we will investigate the phenomenological aspects of the RQM for the massive photon. To do this, we start with 
the equation in (\ref{brqm2}), which describes the motion of the massive photon. Since, for the relativistic massive photon 
satisfying the relation $E^{2}=m^{2}+p^{2}$, we have two kinds of solutions corresponding to $E=\pm (m^{2}+p^{2})^{1/2}\equiv \pm p_{0}$, 
we introduce an ansatz for the wave function $\phi^{a}$ as follows
\beq
\phi^{a}(x)=\phi^{a}(p^{\mu})e^{\mp ip_{\sigma}x^{\sigma}},
\label{phipmu}
\eeq 
for a positive (negative) solution with an upper (lower) sign. 

Next we find the positive energy solution $\phi_{+}^{a}(x)$ with $E>0$ 
and the negative energy solution $\phi_{-}^{a}(x)$ with $E=-|E|<0$, respectively, 
\beq 
\phi_{+}^{a}(x)=u^{a}(p^{\mu})e^{-ip_{\sigma}x^{\sigma}},~~~
\phi_{-}^{a}(x)=v^{a}(p^{\mu})e^{+ip_{\sigma}x^{\sigma}},
\label{phipm}\eeq
where $u^{a}(p^{\mu})$ and $v^{a}(p^{\mu})$ are given by the nontrivial forms,
\beq 
u^{a}(p^{\mu})=\epsilon^{a}\left(\frac{E+m}{2m}\right)^{1/2}\left(\begin{array}{c}
1\\
\frac{p}{E+m}
\end{array}\right),~~~
v^{a}(p^{\mu})=\epsilon^{a}\left(\frac{|E|+m}{2m}\right)^{1/2}\left(\begin{array}{c}
\frac{p}{|E|+m}\\
1
\end{array}\right).
\label{negsol}
\eeq
Here $\epsilon^{a}$ is a unit polarization vector possessing the spacetime index $a$ $(a=0,1,2,3)$ 
which is the same as the spin index and is needed to incorporate minimally the spin DOF for the massive photon. Here we have 
considered the Lorentz frame where $\epsilon^{a}$ is purely space-like so that we can readily find that 
$\epsilon^{a}\epsilon^{a}=\vec{\epsilon}\cdot\vec{\epsilon}=1$, or 
$\epsilon_{a}\epsilon^{a}=-\vec{\epsilon}\cdot\vec{\epsilon}=-1$. Moreover we have the relation $\epsilon_{a}p^{a}\neq 0$, since for the massive photon we have longitudinal component in addition to transverse ones, similar to the phonon associated with massive particle lattice 
vibrations~\cite{phonon}. Next we have the normalization relations: $\bar{u}^{a}u^{a}\equiv u^{a\dagger}\sigma_{3}u^{a}=1$, 
$u^{a~\dagger}u^{a}=\frac{E}{m}$, $\bar{v}^{a}v^{a}\equiv v^{a\dagger}\sigma_{3}v^{a}=-1$ and $v^{a~\dagger}v^{a}=\frac{|E|}{m}$. 
Inserting (\ref{phipm}) into (\ref{brqm2}), we obtain the equations of motion 
in the momentum space
\beq 
(\not\! p - m)u^{a}(p^{\mu})=0,~~~(\not\! p + m)v^{a}(p^{\mu})=0,
\label{pmp} 
\eeq 
where $\not\hspace{-0.135cm}p=\Gamma^{\mu}p_{\mu}$. It is well known that, in the Dirac RQM for the positron, we have the electron 
and positron corresponding to the positive and negative energy solutions, respectively. It is now interesting to note that, similar 
to the Dirac theory for the positron, there exists the massive photon with negative energy solution.

Reshuffling the equation of motion in (\ref{brqm2}) we obtain 
\beq
i\sigma_{3}\pa_{0}\phi^{a}-\sigma_{2}\pa_{3}\phi^{a}-m\phi^{a}=0.
\label{eomsigma1}
\eeq 
Taking Hermitian conjugate of the equation in (\ref{eomsigma1}), we next construct 
\beq
i\pa_{0}\bar{\phi}^{a}\sigma_{3}-\pa_{3}\bar{\phi}^{a}\sigma_{2}+m\bar{\phi}^{a}=0,
\label{hcphi}
\eeq
where 
\beq
\bar{\phi}^{a}\equiv\phi^{a\dagger}\Gamma^{0}.
\label{barphia} 
\eeq
Exploiting (\ref{eomsigma1}) and (\ref{hcphi}), we find the probability 
continuity equation 
\beq
\pa_{0}\rho+\na\cdot\vec{J}=0,
\label{conti}
\eeq
where the probability density $\rho$ and probability current $\vec{J}$, respectively, are given by
\bea
\rho&\equiv&\bar{\phi}^{a}\sigma_{3}\phi^{a}=\phi^{a\dagger}\phi^{a}=\phi_{1}^{*}\phi_{1}+\phi_{2}^{*}\phi_{2},\nn\\
\vec{J}&\equiv&\bar{\phi}^{a}(i\sigma_{2})\phi^{a}\hat{z}=\phi^{a\dagger}\sigma_{1}\phi^{a}\hat{z}=(\phi_{1}^{*}\phi_{2}+\phi_{2}^{*}\phi_{1})\hat{z}.
\label{vectorj}
\eea
Here we have used the notation $\phi^{a}\equiv\epsilon^{a}(\phi_{1},\phi_{2})^{t}$. Note that the probability density $\rho$ is positive definite in our model. We emphasize that the quantities $\rho$ and $\vec{J}$ in (\ref{vectorj}) are physically well defined to yield a good quantization, similar to the Dirac RQM for the electron 
where the corresponding probability density $\rho_{D}$
is also positive definite. However, in the Klein-Gordon model, the probability density $\rho_{KG}$ for a relativistic spinless boson is not positive 
definite~\cite{bjorken64}. Note that the probability continuity equation in (\ref{conti}) can be rewritten in the covariant form as follows
\beq
\pa_{\mu}J^{\mu}=0,
\label{probconti}
\eeq
where we have used the four probability current $J^{\mu}=(\rho,\vec{J})$. 
For the positive energy solution with $E>0$, inserting $\phi^{a}_{+}(x)$ in (\ref{phipm}) into 
$\rho$ and $\vec{J}$ in (\ref{vectorj}), we obtain 
\beq
\rho=\frac{E}{m},~~~\vec{J}=\frac{\vec{p}}{m}.
\label{rhovecj1} 
\eeq
Next, for the negative energy solution with $E=-|E|<0$ where 
$|E|=(m^{2}+p^{2})^{1/2}$, inserting $\phi^{a}_{-}(x)$ in (\ref{phipm}) into $\rho$ and $\vec{J}$, we find
\beq
\rho=\frac{|E|}{m},~~~\vec{J}=\frac{\vec{p}}{m}. 
\label{rhovecj2} 
\eeq

It seems appropriate to comment on the four probability current $J_{D}^{\mu}=(\rho_{D},\vec{J}_{D})$ in the Dirac RQM for an 
electron with mass $m_{e}$. The relativistic equation of motion for the electron is given by~\cite{bjorken64}
\beq
(i\gamma^{\mu}\pa_{\mu}-m_{e})\psi(x)=0,
\label{eomdirac}
\eeq  
with $\gamma^{\mu}$ $(\mu=0, 1, 2, 3)$ being given by $4\times 4$ matrices $\gamma^{0}=\beta$ and $\gamma^{i}=\beta\alpha_{i}$ where
\beq
\alpha_{i}=\left[\begin{array}{cc}
0 &\sigma_{i}\\
\sigma_{i} &0
\end{array}\right],~~~
\beta=\left[\begin{array}{cc}
I &0\\
0 &-I
\end{array}\right].
\label{vecalpha}
\eeq
The electron wave equation is then given by
\beq
\psi(x)=\psi(p^{\mu})e^{\mp ip_{\sigma}x^{\sigma}},
\label{psixdirac}
\eeq
for a positive (negative) solution with an upper (lower) sign. Now we find the positive energy solution $\psi_{+}(x)$ with 
$E_{e}>0$ and the negative energy solution $\psi_{-}(x)$ with $E_{e}=-|E_{e}|<0$, respectively. For a given index $I$ $(I=1, 2)$ 
corresponding to spin $\pm 1/2$ states we construct
\beq 
\psi_{+}^{I}(x)=u^{I}(p^{\mu})e^{-ip_{\sigma}x^{\sigma}},~~~
\psi_{-}^{I}(x)=v^{I}(p^{\mu})e^{+ip_{\sigma}x^{\sigma}},
\label{psipm}\eeq
where, for a given relativistic four momentum variable $p_{e}^{\mu}=(E_{e},\vec{p}_{e})$, $u^{I}(p^{\mu})$ and 
$v^{I}(p^{\mu})$ are given by 
\beq 
u^{I}(p^{\mu})=\left(\frac{E_{e}+m_{e}}{2m_{e}}\right)^{1/2}\left(\begin{array}{c}
\chi^{I}\\
\frac{\vec{\sigma}\cdot\vec{p}_{e}}{E_{e}+m_{e}}\chi^{I}
\end{array}\right),~~~
v^{I}(p^{\mu})=\left(\frac{|E_{e}|+m_{e}}{2m_{e}}\right)^{1/2}\left(\begin{array}{c}
\frac{\vec{\sigma}\cdot\vec{p}_{e}}{|E_{e}|+m_{e}}\chi^{I}\\
\chi^{I}
\end{array}\right).
\label{negsoldirac}
\eeq
where $\chi^{1}=(1,0)^{t}$ and $\chi^{2}=(0,1)^{t}$. In the Dirac RQM for the electron, the four probability current 
$J_{D}^{\mu}=(\rho_{D},\vec{J}_{D})$ is given by~\cite{bjorken64} 
\beq
\rho_{D}=\psi^{\dagger}\psi,~~~J^{i}_{D}=\psi^{\dagger}\alpha_{i}\psi.
\eeq
Making use of (\ref{psixdirac})--(\ref{negsoldirac}), we find 
\beq
\rho_{D}=\frac{2E_{e}}{m_{e}},~~~\vec{J}_{D}=\frac{2\vec{p}_{e}}{m_{e}},
\label{rhovecj1dirac} 
\eeq
for the positive energy solution with $E_{e}>0$, and 
\beq
\rho_{D}=\frac{2|E_{e}|}{m_{e}},~~~\vec{J}_{D}=\frac{2\vec{p}_{e}}{m_{e}},
\label{rhovecj2dirac} 
\eeq
for the negative energy solution with $E_{e}=-|E_{e}|<0$. Here we notice that the four probability current $J^{\mu}_{D}$ satisfies 
the probability continuity equation both for the positive and negative solutions~\cite{bjorken64}
\beq
\pa_{\mu}J^{\mu}_{D}=0.
\label{contidirac}
\eeq
Note that there exists the similarity such as $J^{\mu}$ in 
(\ref{rhovecj1}) and (\ref{rhovecj2}), and $J_{D}^{\mu}$ in (\ref{rhovecj1dirac}) and (\ref{rhovecj2dirac}) 
between the RQM for the massive photon and the Dirac RQM for the electron.

Now we have some comments to address on the negative solution $\phi_{-}^{a}$ in the RQM for the massive photon. 
First, for the negative energy solution of the massive photon, since we have $\rho=|\phi_{1}|^{2}+|\phi_{2}|^{2}>0$, $\rho=\frac{|E|}{m}$ implies that $m$ is 
positive. Even in this negative energy solution case, the positive mass $m$ moves with the probability current $\vec{J}$ along the $z$ direction. 
From now on, we will name the photon possessing the characteristic that the particle has the positive mass $m$ and positive energy $|E|$ and is associated with the negative energy solution, an anti-photon. To be more specific, we find that the anti-photon possessing the positive mass $m$ has a positive definite probability 
density $\rho$ and propagates with a probability current $\vec{J}$ along the direction of $\vec{p}$.

Second, we propose that the {\it anti}-photon related with the negative energy solution is defined to interact repulsively with the ordinary massive photon, {\it oppositely} to the ordinary massive photon-photon attractive gravitational {\it interaction pattern}.\footnote{Note that, in the Dirac RQM, the positron (or {\it anti}-electron) associated with the negative energy solution is defined to interact attractively with the electron, {\it oppositely} to the ordinary charged electron-electron repulsive 
electromagnetic {\it interaction pattern}. The same logic can be applied to the gravitational interaction case related with the anti-photon.} 
Next, the positron and electron can annihilate each other via the 
particle and anti-particle pair annihilation mechanism. However, the uncharged anti-photon scatters away from the ordinary massive particle including the photon, in the repulsive gravitational interaction between the anti-photon and the ordinary massive particle. 

Third, since the anti-photon is repulsive against charged or uncharged ordinary massive matters, the anti-photon does not adhere to the ordinary massive matters, so that the anti-photons can yield an intense radiation flare. Now we propose that the anti-photons with positive masses could be the intense radiation flare of the GRB released from a supernova, during a high-mass star implodes and then forms a neutron star or 
a black hole. 

Next we diagonalize the Hamiltonian in (\ref{hpsi2}) by introducing a unitary operator matrix and 
its inverse one, respectively,
\bea
U&=&\left(\frac{m+(m^{2}-\triangle)^{1/2}}{2(m^{2}-\triangle)^{1/2}}\right)^{1/2}
\left(
\begin{array}{cc}
1 &\frac{i\partial_{3}}{m+(m^{2}-\triangle)^{1/2}}\\
\frac{-i\partial_{3}}{m+(m^{2}-\triangle)^{1/2}} &1
\end{array}
\right),\nn\\
U^{-1}&=&\left(\frac{m+(m^{2}-\triangle)^{1/2}}{2(m^{2}-\triangle)^{1/2}}\right)^{1/2}
\left(
\begin{array}{cc}
1 &\frac{-i\partial_{3}}{m+(m^{2}-\triangle)^{1/2}}\\
\frac{i\partial_{3}}{m+(m^{2}-\triangle)^{1/2}} &1
\end{array}
\right),\nn\\
\label{uni}
\eea
where $\triangle=\partial_{3}\partial_{3}$ is the Laplacian operator. Acting the unitary operator and its 
inverse one on the Hamiltonian $H$ in (\ref{hpsi2}), we construct the diagonal Hamiltonian matrix $H_{D}$ of the form
\beq
H_{D}=U^{-1}HU=\left(\begin{array}{cc}
(m^{2}-\triangle)^{1/2} &0\\
0 &-(m^{2}-\triangle)^{1/2}
\end{array}
\right).
\label{hd}
\eeq  
Here one can readily check that the eigenfunctions for the diagonal Hamiltonian matrix (\ref{hd}) are given by the 
positive and negative energy solutions in (\ref{phipm}).

Finally we give couple of comments on the wave function $\phi^{a}_{A}$ in (\ref{phi2t}). Note that, after some manipulations 
using (\ref{brqm2}), we find the equation of motion in terms of $\phi^{a}_{A}$
\beq
(\Box+m^{2})\phi^{a}_{A}(x)=0.\label{boxm2}
\eeq
Even though the equation of motion in (\ref{boxm2}) seems to be simpler than that in (\ref{brqm2}), in investigating the phenomenology 
of the massive photon, the wave function $\phi^{a}_{A}(x)$ is not so physically meaningful since it is described in terms of the rather non-physical component index $A$, instead of the physical energy index $\pm$ in (\ref{phipm}). The equation of motion 
in (\ref{boxm2}) will be discussed in the next section, to investigate the differences among the RQM for the massless photon, 
the massive Proca model and the RQM for the massive photon.

\section{RQM for massless photons}
\label{antiparticle}
\setcounter{equation}{0}
\renewcommand{\theequation}{\arabic{section}.\arabic{equation}}

In this section, we will formulate  wave functions for massless photons. To do this, we revisit the RQM equation of 
motion in (\ref{brqm2}) and the solutions in (\ref{phipm}) and (\ref{negsol}) from which, in the massless limit, we obtain the following positive
and negative energy solutions 
\beq
\phi_{+}^{a(m=0)}(x)={\cal N}\epsilon^{a}\left(\begin{array}{c}
1\\
1\end{array}\right)
e^{-ip_{\sigma}x^{\sigma}},~~~
\phi_{-}^{a(m=0)}(x)={\cal N}\epsilon^{a}\left(\begin{array}{c}
1\\
1\end{array}\right)
e^{+ip_{\sigma}x^{\sigma}},
\label{photon}
\eeq
where a normalization factor ${\cal N}$ will be fixed later. Note that the above
solutions in (\ref{photon}) satisfy the following equation of motion
\beq
\Box\phi^{a(m=0)}_{\pm}(x)=0,
\label{boxphi}
\eeq 
which can be readily obtained from the massless limit of the RQM equation of motion in (\ref{boxm2}) for the massive photon.

Note also that the positive and negative solutions $\phi^{a}_{\pm}(x)$ in (\ref{phipm}) describing the massive photon and anti-photon, respectively, are given by the nontrivial linear combinations of the wave functions $\phi^{a}_{A=1}$ and $\phi^{a}_{A=2}$. However, in the massless photon, $\phi_{+}^{a(m=0)}(x)$ and $\phi_{-}^{a(m=0)}(x)$ are given by the trivial linear combinations 
of $\phi_{A=1}^{a(m=0)}(x)$ and $\phi_{A=2}^{a(m=0)}(x)$ as shown in (\ref{photon}). Moreover, 
since the neutral massless photon is equal to its massless anti-particle, we construct the massless photon wave 
functions in terms of the wave functions possessing the spin index only. 
In other words, by exploiting $\phi_{\pm}^{a(m=0)}(x)$ in (\ref{photon})
and the unit polarization vector $\epsilon^{a}$, we find $\phi^{a(m=0)}(x)$ corresponding to the massless 
photon wave function possessing the spin index $a$ only without the component index $A$
\beq
\phi^{a(m=0)}(x)=\frac{\epsilon^{a}}{\sqrt{2p_{0}V}}
(e^{-ip_{\sigma}x^{\sigma}}+e^{+ip_{\sigma}x^{\sigma}}). 
\label{munu}
\eeq 
Here the normalization factor ${\cal N}\equiv \frac{1}{\sqrt{2p_{0}V}}$ associated with $p_{0}$ and space volume $V$ is now fixed so that the
massless photon energy $\omega$ in the electromagnetic wave can be given by 
$\omega=p_{0}=|\vec{p}|$~\cite{bjorken64}. Note that $\phi^{a(m=0)}(x)\equiv A^{a}(x)$ is the four-vector potential in the 
elecromagnetism for the massless photon.

It seems appropriate to comment on the normalization factor ${\cal N}$. 
In the Bose-Einstein statistics~\cite{reif}, we can find the massive photon statistics where 
the total number of the photons is countable and fixed so that we can have the constraint of the form $\sum_{r}n_{r}=N$. Here 
$n_{r}$ is the number of photons in quantum state $r$ and $N$ is the total photon number. However in treating the massless photon 
we have a puzzle that, the total number of the massless photons is not fixed so that we cannot have the above constraint on the total number of the photons. Now, this puzzle can be solved by the intrinsic property that the massless photon is a point-like particle without 
any size, differently from the massive photon with finite size.  After quantization of the light we thus cannot count the number of quantized massless photons. Without resorting to the above constraint on the total number of the photons, the quantum statistics for the massless photon is then known to yield the Planck distribution~\cite{reif} $\bar{n}_{s}=\frac{1}{e^{\beta E_{s}}-1}$, where $\beta=1/kT$ with $k$ and $T$ 
being the Boltzmann constant and temperature, respectively, and $E_{s}$ $(s=1,2,...)$ is the energy in state $s$. For the 
massless point-like photons, we thus have the normalization factor in (\ref{munu}), which is different from those in (\ref{negsol}) 
for the massive photon with finite size.

Note that in constructing $\phi^{a(m=0)}(x)$, we have reduced the eight components of the wave function for the massive photon 
into the four components of the wave function for the massless photon. Here we have included only the DOF originated from the spin index 
$a$ in $\phi^{a(m=0)}(x)$ for the massless photon. In other words, we do not have the DOF associated with the 
positive and negative energy solutions, and thus $\phi^{a(m=0)}(x)$ do not have the energy index. 
This construction is consistent with the traditional photon relativistic representation. Note also that the polarization vector $\epsilon^{a}$ satisfies the transversality condition 
$\epsilon_{a}p^{a}=0$, which is needed since the massless photon has the transverse 
components only. Moreover we can readily check that 
the above wavefunction in (\ref{munu}) fulfills the equation of motion for the massless photon 
\beq
\Box \phi^{a(m=0)}(x)=0.
\label{boxaa}
\eeq
The above transversality condition $\epsilon_{a}p^{a}=0$ then yields
$\vec{\epsilon}\cdot\vec{p}=0$, so that we can define two space-like polarization vectors $\vec{\epsilon}_{I}$ $(I=1,2)$ satisfying 
$\vec{\epsilon}_{I}\cdot\vec{\epsilon}_{J}=\delta_{IJ}$. Note that $\vec{\epsilon}_{I}$ and $\vec{p}$ form a three
dimensional orthogonal basis system as desired~\cite{bjorken64}.

Next, in the massive Proca model described by a wave function $\varphi^{a}(x)$ with spin index $a$ only, we find the relativistic 
equation of motion for a massive photon 
\beq
(\Box+m^{2}) \varphi^{a}(x)=0,
\label{boxaa2}
\eeq
which is different from (\ref{boxm2}), since $\varphi^{a}(x)$ in (\ref{boxaa2}) does not possess the component index $A$. 
Here the Hamiltonian is given by a $1\times 1$ matrix. Note that the wave function $\varphi^{a}(x)$ in (\ref{boxaa2}) describes only 
a positive energy solution for the massive photon, and thus it cannot explain the anti-photon aspects 
discussed in the Dirac type RQM for the massive photon.

\section{Lorentz transformation for a massive photon}
\setcounter{equation}{0}
\renewcommand{\theequation}{\arabic{section}.\arabic{equation}}
\label{lorentztrfm}
 
In this section we will investigate the Lorentz transformation for a massive photon. 
To do this, we first consider the infinitesimal Lorentz transformation given by
\beq
x^{\mu\prime}=a^{\mu}_{~\nu}x^{\nu}\equiv (g^{\mu}_{~\nu}+\epsilon^{\mu}_{~\nu})x^{\nu}
\label{amunu}
\eeq
where $\epsilon^{\mu}_{~\nu}$ are the full Lorentz group transformation parameters. Note that for the positive and negative energy 
solutions for a given relativistic four momentum variable $p^{\mu}=(E,\vec{p})$ in the RQM for the 
massive photon, we obtain the four probability current $J^{\mu}$ for finding the photon in (\ref{rhovecj1}) 
and (\ref{rhovecj2}). Next the probability 
density and current in (\ref{conti}) must form a four vector under the Lorentz transformation in order to confirm 
the covariance of the continuity equation and of the probability interpretation associated with the Born's rule in a space-plus-time split 
of spacetime manifold, as in the Dirac RQM for the electron. Moreover the equation for the massive photon in 
(\ref{brqm2}) should be shown to be Lorentz covariant.\footnote{Note that we find the four probability current $J^{\mu}_{D}$ for 
finding the electron in (\ref{rhovecj1dirac}) and (\ref{rhovecj2dirac}), for the positive and negative 
energy solutions with a given relativistic four momentum variable $p_{e}^{\mu}=(E_{e},\vec{p}_{e})$ in the Dirac RQM for the electron. The probability density and current in 
(\ref{contidirac}) should form a four vector under the Lorentz transformation in order to ensure the covariance of the continuity equation 
and of the probability interpretation, and the Dirac equation for the electron in (\ref{eomdirac}) must be shown to be Lorentz covariant. 
These characteristics have been well established in the Dirac RQM for the electron~\cite{bjorken64}. Similar relativistic physics arguments 
can be applied to the massive photon, and one can find that a {\it massive} photon analog of 
the Dirac wave equation in (\ref{psixdirac})--(\ref{negsoldirac}) for a {\it massive} electron exists, as shown 
in (\ref{phipmu})--(\ref{negsol}).}

Exploiting (\ref{amunu}), we will show the Lorentz 
covariance of the relativistic equation of motion for the massive photon in (\ref{brqm2}). Now we introduce 
an equation which takes the form of (\ref{brqm2}) in the primed system
\beq
(i\Gamma^{\mu}\pa_{\mu}^{~\prime}-m)\phi^{a\prime}(x^{\prime})=0.
\label{brqm2prime}
\eeq
Similar to the scheme exploited in the Dirac theory for the electron~\cite{bjorken64}, we now make an ansatz for $\phi^{a\prime}(x^{\prime})$ as follows
\beq
\phi^{a\prime}(x^{\prime})=S(a)\phi^{a}(x).
\label{smatrix}
\eeq
Here $S(a)$ is a function of $a^{\mu\nu}$ and a $2\times 2$ matrix acting on the 2$_{energy}$-component colume vector $\phi^{a}(x)$ for 
a given spin index $a$. In order to find $S(a)$ satisfying (\ref{smatrix}), we manipulate (\ref{brqm2prime}) to yield
\beq
[iS^{-1}(a)\Gamma^{\mu}a_{\mu}^{~\nu}S(a)\pa_{\nu}-m]\phi^{a}(x)=0,
\label{identity}
\eeq
where $S^{-1}(a)$ is an inverse matrix of $S(a)$. Making use of (\ref{brqm2prime}) and (\ref{identity}), we arrive at
\beq
\Gamma^{\mu}a_{\mu}^{~\nu}=S(a)\Gamma^{\nu}S^{-1}(a).
\label{gammamua}
\eeq
Expanding $S(a)$ in powers of $\epsilon^{\mu\nu}$ given by $a^{\mu\nu}$ in (\ref{amunu}) and keeping only the linear term in the 
infinitesimal generators, we make an ansatz
\beq
S=I-\frac{i}{4}\sigma_{\mu\nu}\epsilon^{\mu\nu},
\label{sigmamunu}
\eeq
where $\sigma_{\mu\nu}(=-\sigma_{\nu\mu})$ are $2\times 2$ matrices in the RQM for the massive photon.

Inserting (\ref{amunu}) and (\ref{sigmamunu}) 
into (\ref{gammamua}), we are left with
\beq
\Gamma^{\mu}\epsilon_{\mu}^{~\nu}=\frac{i}{4}[\Gamma^{\nu},\sigma_{\rho\sigma}]\epsilon^{\rho\sigma}
\eeq
from which we obtain the covariance condition of the form
\beq
[\Gamma^{\mu},\sigma_{\rho\sigma}]=2i(g^{\mu}_{~\rho}\Gamma_{\sigma}-g^{\mu}_{~\sigma}\Gamma_{\rho}).
\label{covariancecond}
\eeq
The problem of finding the Lorentz covariance of the relativistic equation of motion for the massive photon 
in (\ref{brqm2}) under the full Lorentz group transformation is now reduced to that of constructing matrices 
$\sigma_{\rho\sigma}$ satisfying (\ref{covariancecond}). 
Similar to the algorithm used in the Dirac theory for the electron~\cite{bjorken64}, the simplest guess to make is an 
anti-symmetric product of two matrices, and we find that 
\beq
\sigma_{\rho\sigma}=\frac{i}{2}[\Gamma_{\rho},\Gamma_{\sigma}]
\label{sigmarhosigma}
\eeq
is the desired matrix which satisfies the covariance condition in (\ref{covariancecond}). Here we have used 
(\ref{gammamu4}) and the ensuing results for $\sigma_{\rho\sigma}$ in the RQM for the massive photon, given by
\beq
\sigma_{03}=-\sigma_{30}=-i\sigma_{1},~~~\sigma_{\rho\sigma}=0,~{\rm otherwise}.
\label{sigma03}
\eeq
We thus prove that (\ref{brqm2prime}) also holds covariantly in the primed system, and we finally show the covariance of the relativistic equation of motion for the massive photon in (\ref{brqm2}) under the full Lorentz group transformation. Note that, in the RQM for the massive photon, the generators of the full Lorentz group 
$K_{i}=\frac{1}{2}\sigma_{0i}$ and $N_{i}=\frac{1}{4}\epsilon_{ijk}\sigma_{jk}$ $(i=1, 2, 3)$ satisfy the commutation relations 
\beq
[K_{i}, K_{j}]=-i\epsilon_{ijk}N_{k},~~~[N_{i}, N_{j}]=i\epsilon_{ijk}N_{k},~~~[N_{i}, K_{j}]=i\epsilon_{ijk}K_{k},
\eeq 
similar to the corresponding relations in the RQM for the electron~\cite{pokorski87}.

Next we consider explicitly the full Lorentz group transformation for a massive photon whose trajectory is a straight line 
along the z axis. To do this, we introduce a rotation around $z$ axis and a boost along $z$ axis, which are physically of interest in the RQM for the massive photon
\beq
\left(\begin{array}{c}
x^{0\prime}\\
x^{1\prime}\\
x^{2\prime}\\
x^{3\prime}\\
\end{array}
\right)
=\left(
\begin{array}{cccc}
1 &0 &0 &0\\
0 &\cos\theta &\sin\theta &0\\
0 &-\sin\theta  &\cos\theta &0\\
0 &0  &0 &1
\end{array}
\right)
\left(\begin{array}{c}
x^{0}\\
x^{1}\\
x^{2}\\
x^{3}\\
\end{array}
\right)
,~~~
\left(\begin{array}{c}
x^{0\prime}\\
x^{1\prime}\\
x^{2\prime}\\
x^{3\prime}\\
\end{array}
\right)
=\left(
\begin{array}{cccc}
\cosh\omega &0 &0 &-\sinh\omega\\
0 &1 &0 &0\\
0 &0  &1&0\\
-\sinh\omega &0  &0 &\cosh\omega
\end{array}
\right)
\left(\begin{array}{c}
x^{0}\\
x^{1}\\
x^{2}\\
x^{3}\\
\end{array}
\right),
\label{rotboot}
\eeq
where $\theta$ and $\omega$ are  the rotation and boost parameters, respectively. Exploiting (\ref{amunu}) and (\ref{rotboot}) we obtain 
the non-vanishing Lorentz transformation parameters $\epsilon_{rot}^{\mu\nu}$ and $\epsilon_{boost}^{\mu\nu}$ for the rotation and boost with 
the conditions $\theta\ll 1$ and $\omega\ll 1$ given by 
\beq
\epsilon_{rot}^{12}=-\epsilon_{rot}^{21}=-\theta,~~~\epsilon_{boost}^{03}=-\epsilon_{boost}^{30}=\omega.
\label{epsilons}
\eeq

Now, by making use of $\sigma_{\mu\nu}$ in (\ref{sigma03}), 
we explicitly find the corresponding matrix $S$ in (\ref{sigmamunu}) for the rotation 
and boost associated with $\epsilon^{\mu\nu}$ in (\ref{epsilons}). First, for the case of the rotation around $z$ axis, we readily 
obtain
\beq
S_{rot}^{z}=I,
\label{srotz}
\eeq
regardless of the non-vanishing rotation parameters $\epsilon_{rot}^{12}=-\epsilon_{rot}^{21}=-\theta$ in (\ref{epsilons}), 
since we have 
$\sigma_{12}=\sigma_{21}=0$ in (\ref{sigma03}). In particular, for the case of a $2\pi$ radian rotation, we still find 
$S_{rot}^{z}=I$ implying a bosonic photon property 
that it takes a rotation of $2\pi$ radian to return $\phi^{a}(x)$ to its original value. 
Note that for the fermionic electron case it takes a $4\pi$ radian rotation to return $\psi(x)$ to its original value~\cite{bjorken64} and 
this characteristic is also discussed in terms of the M\"obius strip structure of the hypersphere manifold in the hypersphere soliton 
model~\cite{hongpre}. Second, for the case of the boost along $z$ axis related with 
the non-vanishing boost parameters $\epsilon_{boost}^{03}=-\epsilon_{boost}^{30}=\omega$, we find 
\beq
S_{boost}^{z}=I-\frac{1}{2}\omega \sigma_{1}.
\label{sboostz}
\eeq

Next we investigate the covariance of the probability continuity equation in (\ref{probconti}) under the full Lorentz group 
transformation. To do this, we first rewrite the four probability current $J^{\mu}$ in (\ref{vectorj}) in terms of $\Gamma^{\mu}$ 
\beq
J^{\mu}=\bar{\phi}^{a}\Gamma^{\mu}\phi^{a}.
\label{jmu2}
\eeq
Exploiting (\ref{barphia}), (\ref{smatrix}) and (\ref{jmu2}), we construct
\beq
J^{\mu\prime}=\bar{\phi}^{a}\Gamma^{0}S^{\dagger}\Gamma^{0}\Gamma^{\mu}S\phi^{a}.
\label{jmuprime}
\eeq
For the rotation around $z$ axis associated with $S_{rot}^{z}=I$, we readily find
\beq
J_{rot}^{\mu\prime}=J_{rot}^{\mu},
\label{jmuprimerot}
\eeq
to produce
\beq
\pa_{\mu}^{~\prime}J_{rot}^{\mu\prime}=a_{\mu}^{~\alpha}\pa_{\alpha}J_{rot}^{\mu}=\pa_{\mu}J_{rot}^{\mu}=0.
\label{probconti2rot}
\eeq
For the boost along $z$ axis related with $S_{boost}^{z}=I-\frac{1}{2}\omega\sigma_{1}$, keeping only the linear term in the 
infinitesimal generators we obtain the non-vanishing components of 
$J_{boost}^{\mu\prime}$
\beq
J_{boost}^{0\prime}=\bar{\phi}^{a}\Gamma^{0}(I-\omega\sigma_{1})\phi^{a},~~~
J_{boost}^{3\prime}=\bar{\phi}^{a}\Gamma^{3}(I-\omega\sigma_{1})\phi^{a}.
\label{jmuprimeboost}
\eeq
Note that, from (\ref{jmuprimerot}) and (\ref{jmuprimeboost}), the probability density $\rho$ transforms like the time component of the four probability current $J^{\mu}=(\rho,\vec{J})$ of a rank one tensor. Exploiting (\ref{jmuprimeboost}) we obtain
\beq
\pa_{\mu}^{~\prime}J_{boost}^{\mu\prime}=a_{\mu}^{~\alpha}\pa_{\alpha}J_{boost}^{\mu\prime}=(1-\omega^{2})\pa_{\mu}J_{boost}^{\mu}=0.
\label{probconti2boost}
\eeq
Making use of (\ref{probconti2rot}) and (\ref{probconti2boost}), we thus show that in the primed system
\beq
\pa_{\mu}^{~\prime}J^{\mu\prime}=0,
\label{probconti2}
\eeq
to confirm the covariance of the probability continuity equation in (\ref{probconti}) under the full Lorentz group transformation.

\section{Conclusions}
\setcounter{equation}{0}
\renewcommand{\theequation}{\arabic{section}.\arabic{equation}}

In summary, making use of the RQM for the electron, Dirac predicted the existence of the positron before the quantum field theory had not been developed. Similarly, exploiting the RQM for the massive photon, we have predicted the existence of the anti-photon without resorting to the quantum field theory. To do this, following the Dirac formalism for the RQM for the electron, we have developed a physics algorithm for the RQM for a massive photon, to formulate a 2$\times$2 Hamiltonian matrix 
for the massive photon. Exploiting the Hamiltonian, we have found a new theoretical particle solution which allows the massive 
photon possessing either positive or negative energy solutions. In particular, we have proposed theoretically the anti-photon corresponding to 
the negative energy solution, similar to the positron in the Dirac RQM for an electron. In our model, 
the massive photon has been shown to possess the longitudinal polarization in addition to the transverse ones as in the case of a phonon. 
Moreover we have formulated a nontrivial diagonal Hamiltonian for the massive photon. 
We also have investigated the Lorentz transformation associated with the RQM for the massive photon, to ensure 
the covariances of the relativistic equation of motion and the corresponding probability continuity equation. 
One of the main points of this paper is that, the anti-photons could be the candidate for the {\it intense} 
radiation flare of the mysterious GRB released from a supernova which is {\it located 
far from the Earth}. Note that the RQM for the anti-photon possessing positive mass 
is a new phenomenology, which could be consistent with the well-established Dirac positron theory and related with a fundamental prediction of the GRB. 

\acknowledgments{The author would like to thank the anonymous referee for helpful comments. He 
was supported by Basic Science Research Program through 
the National Research Foundation of Korea funded by the Ministry of Education, NRF-2019R1I1A1A01058449.}

\end{document}